\documentclass[aps,amsmath,prl,superscriptaddress,a4paper,floatfix,onecolumn]{revtex4}
\usepackage{graphicx}
\usepackage{amsmath}
\usepackage{amsfonts}
\usepackage{amssymb}

\newcommand{\beq}{\begin{equation}}
\newcommand{\eneq}{\end{equation}}
\newcommand{\bea}{\begin{eqnarray}}
\newcommand{\enea}{\end{eqnarray}}

\begin{document}
\title{Macroscopic quantum tunneling in spin filter ferromagnetic Josephson junctions}

\author{D. Massarotti}
\email{dmassarotti@na.infn.it}
\affiliation{Universit\'{a} degli Studi di Napoli "Federico II", Dipartimento di Fisica, via Cinthia, 80126 Napoli (Na) Italy}
\affiliation{CNR-SPIN UOS Napoli, Complesso Universitario Monte Sant'Angelo via Cinthia, 80126 Napoli (Na) Italy}
\author{A. Pal}
\affiliation{Department of Materials Science and Metallurgy, University of Cambridge, 27 Charles Babbage Road, Cambridge, CB3 0FS, United Kingdom}
\author{G. Rotoli}
\affiliation{Seconda Universit\'{a} degli Studi di Napoli, Dipartimento di Ingegneria Industriale e dell'Informazione, via Roma 29, 81031 Aversa (Ce) Italy}
\author{L. Longobardi}
\affiliation{Seconda Universit\'{a} degli Studi di Napoli, Dipartimento di Ingegneria Industriale e dell'Informazione, via Roma 29, 81031 Aversa (Ce) Italy}
\affiliation{American Physical Society, 1 Research Road, Ridge, New York 11961, USA}
\author{M. G. Blamire}
\affiliation{Department of Materials Science and Metallurgy, University of Cambridge, 27 Charles Babbage Road, Cambridge, CB3 0FS, United Kingdom}
\author{F. Tafuri}
\email{tafuri@na.infn.it}
\affiliation{CNR-SPIN UOS Napoli, Complesso Universitario Monte Sant'Angelo via Cinthia, 80126 Napoli (Na) Italy}
\affiliation{Seconda Universit\'{a} degli Studi di Napoli, Dipartimento di Ingegneria Industriale e dell'Informazione, via Roma 29, 81031 Aversa (Ce) Italy}

\date{\today}

\begin{abstract}

The interfacial coupling of two materials with different ordered phases, such as a superconductor (S) and a ferromagnet (F) is driving new fundamental physics and innovative applications. For example, the creation of spin-filter Josephson junctions and the demonstration of triplet supercurrents have suggested the potential of a dissipationless version of spintronics based on unconventional superconductivity. Here we demonstrate evidence for active quantum applications of S-F-S junctions, through the observation of macroscopic quantum tunneling in Josephson junctions with GdN ferromagnetic insulator barriers. We prove a clear transition from thermal to quantum regime at a crossover temperature of about 100 mK at zero magnetic field in junctions which demonstrate a clear signature of unconventional superconductivity. Following previous demonstration of passive S-F-S phase shifters in a phase qubit, our result paves the way to the active use of spin filter Josephson systems in quantum hybrid circuits.

\end{abstract}


\maketitle

{\bfseries{Introduction}}

Superconductor (S)/Ferromagnet (F) interfaces are a special class of hybrid systems where different ordered phases meet and generate new nanoscale phenomena and new forms of global order\cite{buzdin_rew,efetov_rew}. A ferromagnetic barrier in a Josephson junction (JJ) generates novel physics\cite{ryazanov,robinson}, and represents a technological key for advances in weak superconductivity, spintronics and quantum computation\cite{golubov,Buzdin,kawabata}. Recent interest in ultra-low-power, high-density cryogenic memories has spurred new efforts to simultaneously exploit superconducting and magnetic properties so as to create novel switching elements having these two competing orders\cite{baek}. S-F-S junctions are expected to shed light on several aspects of unconventional superconductivity, including transport through spin aligned triplet Cooper pairs\cite{bergeret_prl,fominov,robinson,robinson3,birge,sprungmann,aarts}.
 
The NbN-GdN-NbN junctions investigated in these experiments are spin filter devices\cite{blamire} with an unconventional predominant second harmonic current-phase relation (CPR)\cite{pal1}. By changing the thickness of the GdN ferromagnetic insulator (FI) barrier it is possible to change its magnetic properties and hence the spin filter efficiency (SFE)\cite{pal1,pal2}. In this paper we report measurements on NbN-GdN-NbN junctions for very low and very high SFE values, from almost a few percent up to 90$\%$, with an intermediate value of 60$\%$. By increasing the thickness of GdN we also obtain junctions with lower values of the \textit{I}$_c$\textit{R}$_n$ product, where \textit{I}$_c$ is the critical current and \textit{R}$_n$ the normal state resistance. Following the previous demonstration by Feofanov et al.\cite{ustinov} of passive S-F-S phase shifters in a phase qubit, the experiment reported here demonstrates the active quantum potential of S-F-S JJs via the occurrence of macroscopic quantum tunneling (MQT) in spin filter devices.

\newpage

{\bfseries{Results}}

{\bfseries{Transport properties of spin filter Josephson junctions}}.

The hallmark of the spin-filter effect is the decrease in resistance in the \textit{R} vs \textit{T} curves below the ferromagnetic transition, as shown in Fig. 1a (\textit{T}$_{Curie} \simeq$ 33 K for GdN). SFE is defined as the percentage difference in the tunnelling probability for up/down spin electrons owing to the difference in barrier heights of the up/down spin channel in the FI caused by exchange splitting, so that 100\% SFE corresponds to pure tunnelling of one spin sign. See Methods for details concerning the calculation of SFE. Table 1 collects the parameters of the measured junctions. All the NbN-GdN-NbN JJs present hysteresis larger than 90$\%$ in the current-voltage (IV) characteristics, in a wide range of Josephson coupling energies \textit{E}$_J=$ \textit{I}$_{c0}$\textit{$\phi$}$_0/2$\textit{$\pi$} (where \textit{I}$_{c0}$ is the critical current in absence of thermal fluctuations and \textit{$\phi$}$_0=$ \textit{h}/2\textit{e} is the quantum flux). For all the measured devices the junction area is about 7$\mu$m $\times$ 7$\mu$m $\simeq$ 50 $\mu$m$^2$.


In Fig. 1d, IV curves, measured as a function of temperature \textit{T} for the junction with the highest SFE, are reported. The dependence of \textit{I}$_c$ as a function of \textit{H} at \textit{T} = 4 K is reported in Fig.  1c.  The blue curve is the first measurement of the magnetic pattern, after nominal zero-field cooling. \textit{I}$_c$(\textit{H}) has then been measured both in the downward direction of the magnetic field sweep and in the upward direction (black and red curves, respectively).  The black and red curves show a distinctive shift of the absolute maximum of \textit{I}$_c$ from $-1$\,mT to $+1$\,mT respectively arising from the hysteretic reversal of the FI barrier\cite{blamire_sust}. The period of \textit{I}$_c$(\textit{H}) in non-spin filter junctions for the same geometry is twice as large (3 mT), pointing to a largely predominant second harmonic in the CPR in spin filter JJs, as discussed in Ref.\cite{pal1}. 

{\bfseries{Measurements of switching current distributions}}.

We have studied the escape rate of the superconducting phase \textit{$\varphi$} as a function of temperature and for different values of the magnetic field, through standard switching current distribution (SCD) measurements\cite{devoret1985,martinis1987,bauch_2005,luigi2}. SCDs have been performed for all the samples reported in Table \ref{table}. According to the Resistively and Capacitively Shunted Junction (RCSJ) model\cite{barone,likharev}, for a JJ with a conventional CPR the dynamics is equivalent to that of a particle of mass \textit{m}$_{\varphi}$ moving in a washboard potential $U(\varphi)= -E_{J} (\cos\!\varphi + \gamma \varphi)$. The particle mass is given by $m_{\varphi}=C(\phi_{0}/2\pi)^{2}$ with \textit{C} the capacitance of the junction. The normalized bias current \textit{$\gamma$}$=$\textit{I}/\textit{I}$_{c0}$ determines the tilt of the potential.

The second harmonic component in the CPR, $I=I_1 \sin\varphi + I_2\sin2\varphi$, leads to a modified washboard potential $U(\varphi)=-E_{1} (\cos\!\varphi + \frac{g}{2} \cos\!2\varphi + \frac{I_{c0}}{I_1} \gamma\varphi)$, \textit{E}$_1=$\textit{$\hbar$}\textit{I}$_1$/2\textit{e}, which may assume the form of a double well for values of \textit{g}$=$\textit{I}$_2/$\textit{I}$_1$ larger than 0.5\cite{tza03, goldobin1,goldobin2,goldobin3} (see Fig. 2a). The presence of two wells in the washboard potential may result in the observation of two critical currents in the IV characteristics, since when tilting back the washboard potential the phase particle may be retrapped in one of the potential wells with finite probability\cite{goldobin1}; the case of "$\varphi$ JJs" with \textit{g} $<-1/2 $ has been recently studied\cite{goldobin2,goldobin3}. Measurements of two-well distinguished critical currents constitute a very direct criterion to estimate the \textit{g} factor\cite{goldobin1,goldobin2}. Although spin filter JJs have a strong second harmonic component in the CPR\cite{pal1} we have not found evidence of two critical currents, and hence the case of $\varphi$ junction with negative values of \textit{g} is not considered in this work.

The $\sin\! 2\varphi$ term in the CPR on average lowers the barrier height of the washboard potential without significantly altering the asymptotic expression of the potential barrier for \textit{$\gamma$} close to one. The height of the potential barrier is given by: $\Delta U (\gamma, g) = \frac{4}{3} E_{J} a(g) \left(1-\gamma \right)^{3/2}$, where $a (g)^2=\frac{\sqrt{32g^2+1}+3}{2\sqrt{32g^2+1}}$, and the plasma frequency is given by: $\omega_p (\gamma, g)= \sqrt[4]{2/a(g)^2} \sqrt{2eI_{c0}/\hbar C} (1-\gamma^2)^{1/4}$. In addition, the second harmonic component does not modify the power law of \textit{$\Delta U$}. As a consequence in the thermal regime the standard deviation \textit{$\sigma$} of the SCDs is expected to scale as \textit{T}$^{2/3}$, as in the case of the standard CPR\cite{devoret1985,martinis1987}.

The motion of the particle is subject to damping given by \textit{Q}$^{-1}$, where $Q=\omega_{p}RC$ is the quality factor and \textit{R} is the resistance of the junction.
When the bias current is ramped from \textit{$\gamma$} $ = 0$ to \textit{$\gamma$} $ <1$, the junction is in the zero voltage state in absence of thermal and quantum fluctuations, corresponding to the particle being localized in a potential well. At finite temperature the junction may switch into the finite voltage state for a bias current \textit{$\gamma$} $ <1$. This corresponds to the particle escaping from the well either by thermally activated processes or by tunneling through the barrier potential (see Fig. 2b). In the thermal activation (TA) regime, the escape rate for weak to moderate damping (\textit{Q} $>1$) is determined by\cite{kramers} $ \Gamma_{t}= a_{t} \frac{\omega_{p}}{2\pi} \exp\left(-\frac{\Delta U}{k_{B}T}\right) $, where the thermal prefactor is $a_{t}=4\left[\left(1+Qk_{B}T/1.8\Delta U \right)^{1/2}+1\right]^{-2}$ \cite{but83}. The escape rate will be dominated by MQT at low enough temperatures\cite{cal81}. For \textit{Q} $ > 1$ and \textit{$\gamma$} close to 1 the escape rate in the quantum regime is: $ \Gamma_{q}= a_{q} \frac{\omega_{p}}{2\pi} \exp\left(-\frac{\Delta U }{\hbar \omega_p}\right) $, where $a_{q}=\left(864\pi \Delta U /\hbar \omega_p \right)^{1/2}$. The crossover temperature between the thermal and quantum regimes is given by\cite{grabert} 

\begin{equation}
	T_{cross}=\frac{\hbar\omega_{p}}{2\pi k_{B}}\left\{\left[1+\left(\frac{1}{2Q}\right)^{2}\right]^{1/2}-\frac{1}{2Q}\right\}\: .
	\label{Tcr}
\end{equation}

The experimental probability density of switching is related to the escape rate through the following equation\cite{fulton}:

\begin{equation}
	P(I)=\frac{\Gamma (I)}{\Delta I/\Delta t} \exp \left[- \int_{0}^{I} {\frac{\Gamma (I')}{\Delta I'/\Delta t}dI'} \right]
      \label{fulton}
\end{equation}

where \textit{$\Delta$I}/\textit{$\Delta$t} is the current ramp rate.

The measurements have been performed in a dilution refrigerator with reaches a base temperature of 20\,mK. A full description of the apparatus is given in detail in the Methods section. The bias current of the junction is ramped at a constant sweep rate \textit{$\Delta$I}/\textit{$\Delta$t} $=2$\,mA s$^{-1}$ and at least 10$^4$ switching events have been recorded using a standard technique\cite{prbLuigi}.

Fig. 3a shows a set of SCDs as a function of temperature for the high SFE JJ reported in Fig. 1. The thermal behavior of the SCDs is typical of underdamped JJs and the standard deviation \textit{$\sigma$}, which is proportional to the width of the switching histograms, increasing with temperature as expected. Fig. 3b shows the SCDs measured below 1\,K (black circles). The dependence of the standard deviation \textit{$\sigma$} on temperature is reported in Fig. 4a (right axis), along with the thermal behavior of the mean value of the SCDs, \textit{I}$_{mean}$, below 0.5\,K (left axis). When decreasing the temperature, \textit{I}$_{mean}$ increases while \textit{$\sigma$} decreases and both saturate at a crossover temperature of about 100\,mK. Below this crossover the histograms overlap and the escape process is no longer regulated by thermal fluctuations, indicating the transition to the MQT regime\cite{devoret1985,martinis1987}.

A further confirmation of the observation of MQT comes from measurements of SCDs in magnetic field. The behavior of \textit{$\sigma$}(\textit{T}) at \textit{H} $ = 1.1$\,mT is shown in Fig. 4b. At this value of the magnetic field, which reduces \textit{I}$_c$ to half of the value measured at zero field (see the blue squares in Fig. 1c), lower values of \textit{$\sigma$} have been measured and \textit{T}$_{cross}$ is reduced by a factor $\sqrt{2}$, down to about 70\,mK, in agreement with MQT theory\cite{grabert}. In both cases of 0\,mT and 1.1\,mT, \textit{T}$_{cross}$ has been determined by the intersection of the \textit{T}$^{2/3}$ curve in the TA regime (dashed green lines in Fig. 4b) and the mean value of \textit{$\sigma$} in the MQT regime (black full lines in Fig. 4b). The measurements in presence of magnetic field prove that the flattening of \textit{$\sigma$} at \textit{H} $ =0$\,mT is a quantum effect and is not due to noise or heating in the measurement setup\cite{devoret1985,martinis1987,bauch_2005,luigi2}.

\newpage

{\bfseries{Discussion}}

In the literature there are no measurements of SCDs on junctions with a dominant second harmonic component in the CPR. Numerical simulations of the phase dynamics as a function of the damping parameter \textit{Q}, the \textit{g} factor and the temperature \textit{T} give the conditions for which a double well potential effectively behaves in the escape process as a single well (for details see Supplementary Fig. 1, Supplementary Fig. 2 and Supplementary Note 1). Namely, for values of \textit{Q} $\simeq 10$ SCDs with a single peak have been obtained for \textit{g} $ \le 2$ or in the limiting case of pure second harmonic CPR (\textit{I}$_1 =0$). In this case the washboard potential changes its periodicity but assumes the form of a single well potential. Instead, for \textit{g} $ \ge 2$, two critical currents should be observed. In fact the heights of the two barriers approach each other when increasing the \textit{g} factor, and the phase may be retrapped in both the potential wells with a finite probability, resulting in a bimodal switching distribution when counting many escape events\cite{goldobin1,goldobin2,goldobin3} (see Supplementary Fig. 3 and Supplementary Discussion). Since this is not observed, a pure second harmonic is the only possible explanation consistent with both measurements of magnetic field pattern\cite{pal1} (see Fig. 1c) and SCDs.

\textit{I}$_{c0}$ can be obtained by fitting the probability density of switching \textit{P}(\textit{I}) (red lines in Fig. 3b) in the thermal and quantum regime, \textit{I}$_{c0}=30.41\pm 0.05$\,$\mu$A. A quite accurate value of the capacitance \textit{C} can be obtained from the crossover temperature \textit{T}$_{cross}$, which depends on \textit{Q}, \textit{C} and \textit{I}$_{c0}$, see equation (1). By inserting the values of \textit{I}$_{c0}$ and \textit{Q} in the expression for \textit{T}$_{cross}$ we get \textit{C}$=4.5\pm 0.9$\,pF. These values lead to \textit{$\omega$}$_{p} \approx 14$\,GHz. Nevertheless, as shown in Fig. 3c, the function \textit{a}(\textit{g}) in the equations for \textit{$\Delta U$} and \textit{$\omega$}$_p$ is a slowly varying function for \textit{g} $>>1$, thus the junction parameters weakly depend on the \textit{g} factor for high values of \textit{g}.

We expect in future additional insights coming from a comparative analysis with samples with lower SFE. For junction cross sections of about 50 $\mu$m$^2$ the \textit{I}$_c$ values for such junctions, are too high to be in the conditions to observe unambiguously the transition from the thermal to the quantum regime as commonly occurring also in standard S-Insulator-S junctions\cite{devoret1985,martinis1987,webb3,han,martinis2002,Berkley}. Only further advances in fabrication able to insert S-FI-S junctions in cavities and qubit architectures will give more refined feedback on the modes of the dissipative domains of the junction, and on the triplet component. However we can infer that the \textit{Q} values of spin filter junctions are relatively higher than one would naively expect on the basis of the properties of the low SFE samples, which are characterized by higher values of \textit{I}$_c$. Concerning the \textit{Q} values of spin filter JJs, the reduction in \textit{I}$_c$ is compensated by the increase of \textit{R}$_n$, as reported in Table 1.

In conclusion, we have demonstrated the occurrence of MQT in NbN-GdN-NbN spin filter JJs. Spin filtering drives the S-F-S junction in the underdamped regime and in the appropriate window of junction parameters to observe MQT. The switching current distributions, together with the period of \textit{I}$_c$(\textit{H}) modulation provides direct evidence for a pure second harmonic current-phase relation in the junction where MQT was observed. This is clear evidence of unconventional superconductivity, and it is possible that transport occurs by means of a pair of spin aligned triplet Cooper pairs\cite{pal1} which may suppress magnetic sources of decoherence\cite{houzet,trifunovic}. Demonstration of macroscopic quantum phenomena in spin filter devices gives promise for their application in quantum hybrid circuits\cite{ustinov}, and also possibly as quiet memories.

\newpage

{\bfseries{Methods}}


{\bfseries{Determination of spin filter efficiency}}.

Spin filter efficiency (SFE) at a particular temperature is calculated from the \textit{R} vs \textit{T} curve. SFE at any temperature below the Curie temperature (\textit{T}$_{Curie}$) of GdN and above the superconducting transition temperature (\textit{T}$_c$ ) of NbN is calculated by defining SFE $=\frac{\sigma \uparrow - \sigma \downarrow}{\sigma \uparrow + \sigma \downarrow}$ where \textit{$\sigma$}$\uparrow$, \textit{$\sigma$}$\downarrow$ are current densities of the up and down spin channels respectively. By invoking WKB approximation, we can derive SFE $= \tanh \left( \cosh^{-1} \left( R^{*}/R \right) \right)$ where \textit{R}$^*$ is the measured value of resistance, and  \textit{R} is the value of resistance in the absence of spin filtering. \textit{R} is estimated by fitting an exponential to the \textit{R} vs \textit{T} curve to temperatures above 50 K, and extrapolating the exponential to temperatures below the \textit{T}$_{Curie}$ of GdN.
For calculating SFE below \textit{T}$_c$ of NbN, an exponential is fitted to the temperature dependence of SFE between \textit{T}$_{Curie}$ and \textit{T}$_c$ and the fitted exponential is extrapolated to lower temperatures to obtain an estimate of SFE.

{\bfseries{Setup for switching current distributions}}.

The SCDs have been measured by thermally anchoring the samples to the mixing chamber of He$^3$/He$^4$ Oxford dilution refrigerator. The bias current is ramped at a constant sweep rate \textit{$\Delta$I}/\textit{$\Delta$t} of about 2 mA s$^{-1}$, the voltage is measured using a low noise differential amplifier and is fed into a threshold detector which is set to generate a pulse signal when the junction switches from the superconducting state to the finite  voltage state. This signal is used to trigger a fast volt meter to record the value of the switching current. This procedure is repeated at least $10^4$ times at each temperature, which allows to compile a histogram of the switching currents. Filtering is guaranteed by  a room temperature electromagnetic interference filter stage followed by low pass RC filters with a cut-off frequency of 1.6 MHz anchored at 1.5 K, and by a combination of copper powder and twisted pair filters thermally anchored at the mixing chamber of the dilution refrigerator.

{\bfseries{Acknowledgments}}


We acknowledge financial support from COST Action MP1201 [NanoSC COST], by Progetto FIRB HybridNanoDev RBFR1236VV001 and by Regione Campania through POR Campania FSE 2007/2013, progetto MASTRI CUP B25B09000010007.



{\bfseries{Author Contributions}}

D.M., A.P., L.L., M.G.B. and F.T. conceived the experiments, A.P. and M.G.B. designed and realized the junctions; D.M. carried out the measurements; D.M., G.R. and F.T. worked on the theoretical modelling and data analysis; D.M., M.G.B. and F.T. co-wrote the paper. All authors discussed the results and commented on the manuscript.

{\bfseries{Competing financial interests}}

The authors declare no competing financial interests.


\newpage

\begin{table}
\caption{\label{table} {\bfseries{Device parameters.}} Spin filter efficiency has been determined at 4.2 K (see Methods for details). \textit{I}$_{c0}$ has been estimated from the fits of the switching current distributions and \textit{R}$_n$ from the IV curves.}
\begin{tabular}{ccccccc}
Barrier thickness (nm) & Spin filter efficiency &  \textit{I}$_{c0}$ ($\mu$A) & \textit{E}$_J$ (meV) & \textit{I}$_c$\textit{R}$_n$ (mV) & \\ \hline \\
1.5 &  $<$30\% & 820  & 1700 & 1.0     \\
1.7 &  $<$30\%  &  280 &  570  & 0.9   \\
1.8  & 60\% &  120 & 250  & 0.7  \\
3.0 & 90\% & 30 & 60 & 0.1 \\
\end{tabular}
\end{table}

\begin{figure}
\includegraphics[width=\linewidth]{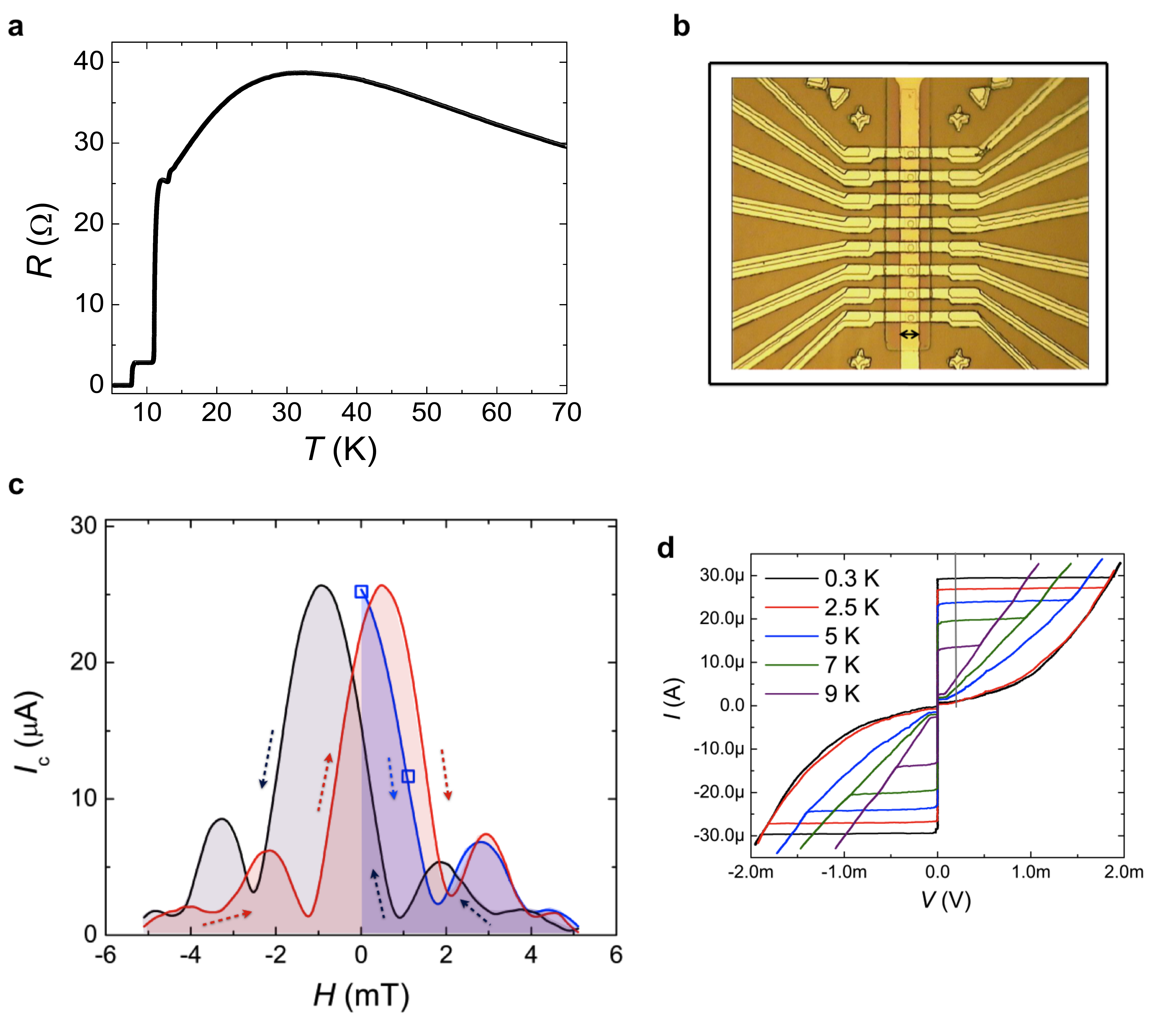}
\caption{{\bfseries{Spin filter Josephson junctions.}} ({\bfseries{a}}) Thermal behavior of the resistance is a hallmark of spin filter NbN-GdN-NbN Josephson junctions (JJs). Semiconducting behavior has been observed at high temperatures, while below the Curie temperature of GdN ($\simeq$ 33 K) the resistance decreases when decreasing the temperature. In non spin filter NbN-GdN-NbN JJs a semiconducting behavior has been observed down to the transition temperature of NbN. In panel {\bfseries{b}} a picture of the device area is shown. The black arrow is the scale bar, 10\,$\mu$m. ({\bfseries{c}}) Magnetic field pattern of a spin filter JJ in the range [-5\,mT, 5\,mT], measured at 4 K. The blue curve is measured after nominal zero field cooling, the black and red curves are the magnetic pattern in the downward and upward direction of the magnetic field sweep, respectively, as indicated by the coloured dashed arrows. Blue squares indicate the values of the magnetic field, for which we present the measurements of switching current distributions. ({\bfseries{d}}) Current-voltage characteristics of the same junction are shown as a function of the temperature, along with the voltage threshold (gray line) used for the switching current measurements.}
\label{pattern}
\end{figure}

\begin{figure}
\includegraphics[width=0.8\linewidth]{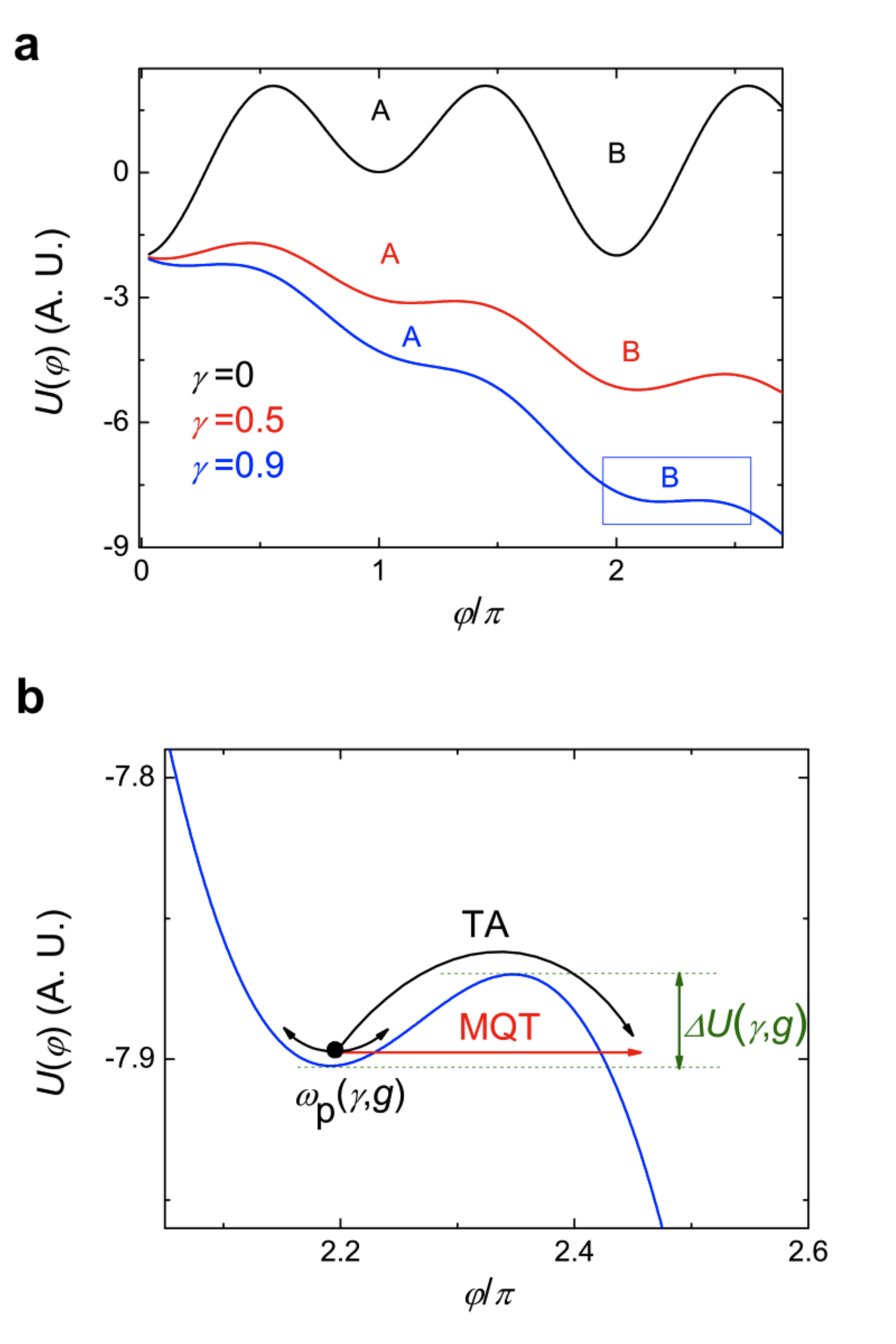}
\caption{{\bfseries{Phase dynamics with strong second harmonic component.}} Washboard potential in the case of a strong second harmonic component in the current-phase relation, for different values of the bias current \textit{$\gamma$}. Both panels {\bfseries{a}} and {\bfseries{b}} refer to the case \textit{g} $=2$, for which two potential wells are present. A and B indicate the lower and the higher potential well, respectively. The blue rectangle in {\bfseries{a}} is magnified in panel {\bfseries{b}}, in which thermal activation (black arrow) and quantum tunneling processes (red arrow) from the well B at \textit{$\gamma$} $=0.9$ are shown. \textit{$\omega$}$_p$ (\textit{$\gamma$}, \textit{g}) and \textit{$\Delta U$} (\textit{$\gamma$}, \textit{g}) are the plasma frequency and the height of the potential barrier respectively.}
\label{washboard}
\end{figure}

\begin{figure}
\includegraphics[width=0.5\linewidth]{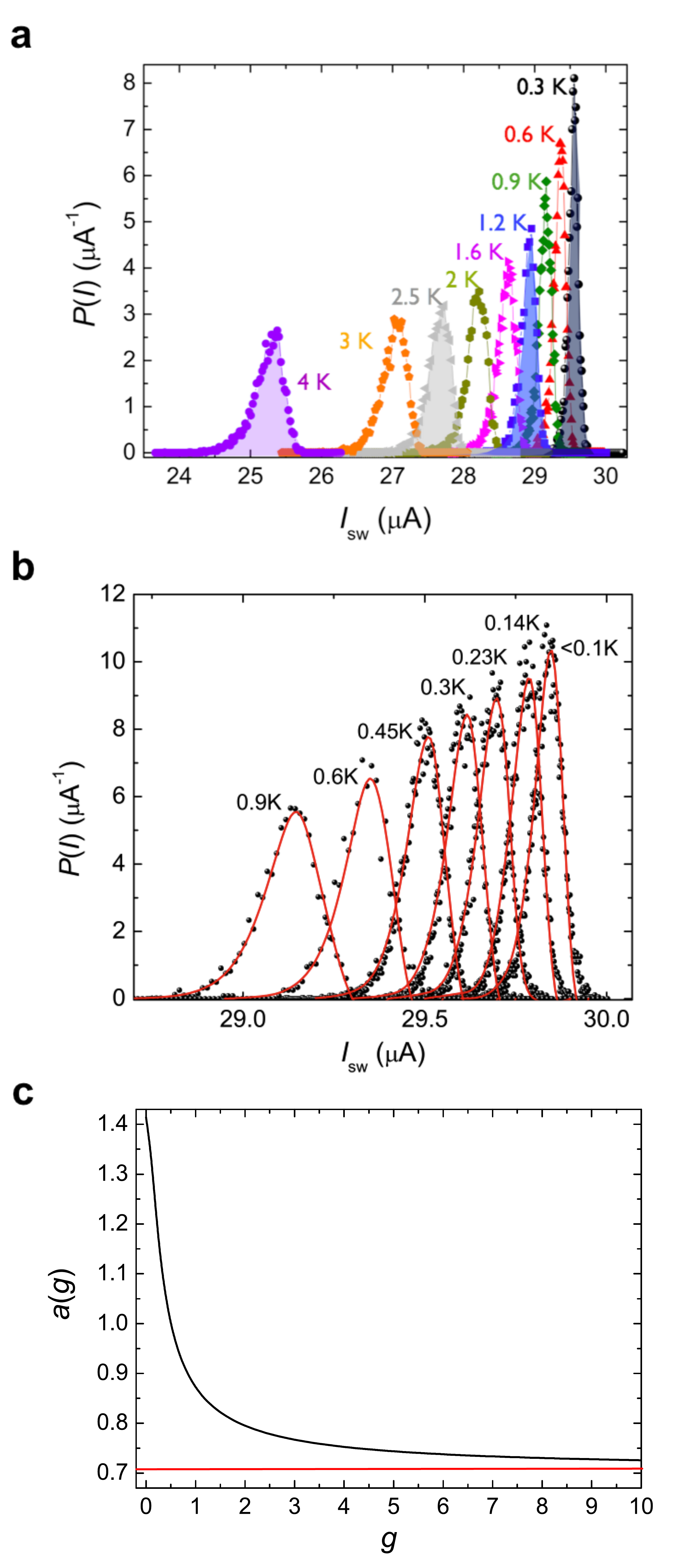}
\caption{{\bfseries{Switching current distributions.}} ({\bfseries{a}}) Measurements of switching current distributions (SCDs) from 4\,K down to 0.3\,K. When increasing the temperature, the histograms move to lower values of switching currents \textit{I}$_{sw}$ and the standard deviation \textit{$\sigma$} increases. Each histogram collects 10$^4$ switching events. ({\bfseries{b}}) Thermal behavior of the SCDs from 0.9\,K down to 20\,mK. The red lines are fits of the probability density of switching, according to equation (2) in the case of a dominant second harmonic component in the current-phase relation (CPR). Below 100\,mK the histograms overlap, indicating the transition to the quantum regime, definitely confirmed by the their dependence in magnetic field. In panel {\bfseries{c}} the function \textit{a}(\textit{g}) is plotted (black curve) and the red line is the value of the function in the limiting case of a pure second harmonic (CPR).}
\label{histograms}
\end{figure}

\begin{figure}
\includegraphics[width=0.8\linewidth]{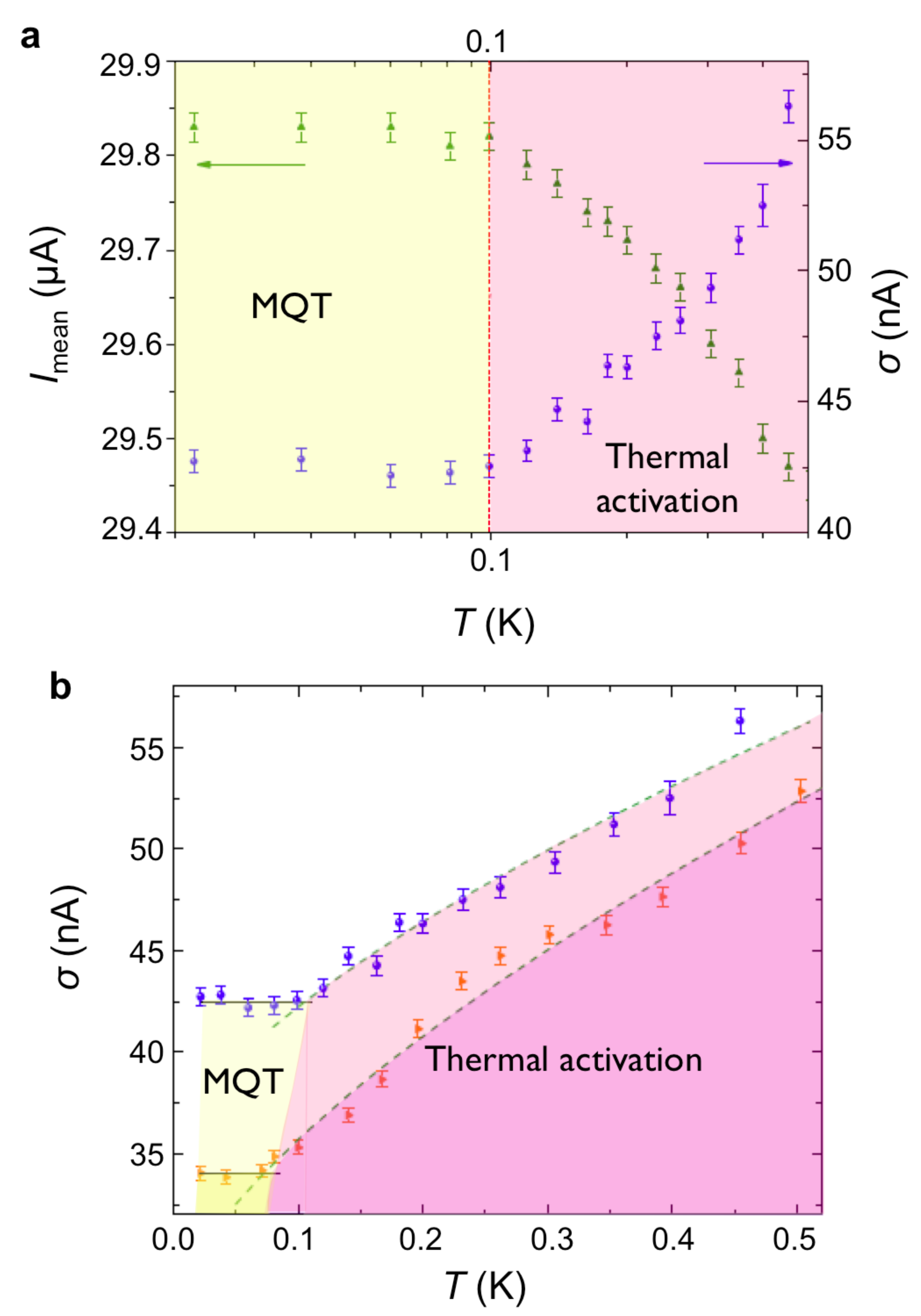}
\caption{{\bfseries{MQT in spin filter S-F-S Josephson junctions.}} ({\bfseries{a}}) Thermal behavior of the mean switching current \textit{I}$_{mean}$ (green up triangles, left axis) and of the standard deviation \textit{$\sigma$} (violet circles, right axis) in absence of magnetic field. The red dashed line indicates the crossover temperature between macroscopic quantum tunneling (MQT) regime (dark yellow background) and thermal activation (pink background). In the MQT regime both \textit{I}$_{mean}$ and \textit{$\sigma$} saturate. ({\bfseries{b}}) Comparison between thermal dependences of \textit{$\sigma$} at 0\,mT (violet circles) and at 1.1\,mT (orange right triangles). In presence of magnetic field, lower values of \textit{$\sigma$} have been measured and \textit{T}$_{cross}$ is reduced, according to equation (1). \textit{T}$_{cross}$ has been determined by the intersection of the \textit{T}$^{2/3}$ curve in the thermal activation regime (dashed green lines) and the mean value of \textit{$\sigma$} in the MQT regime (black full lines). In ({\bfseries{a}}) and in ({\bfseries{b}}) the error bars have been determined by using the theory for propagation of statistical errors, from the equations defining the mean and the standard deviation of the switching histograms.}
\label{sigma}
\end{figure}








\end{document}